\documentclass[aps,prd,showpacs,showkeys,preprintnumbers,amsmath,amssymb,nofootinbib,11pt]{revtex4}
\usepackage{eepic}
\usepackage{indentfirst}
\usepackage{mathrsfs}
\usepackage{fancyhdr}
 \usepackage{ulem}
\usepackage{tabularx}
\usepackage{graphicx}
\usepackage{subfigure}
\usepackage{epsfig}
\usepackage{color}
\usepackage{slashed}
\newcommand{\be}{\begin{equation}}
\newcommand{\ee}{\end{equation}}
\newcommand{\ba}{\begin{array}{c}}
\newcommand{\ea}{\end{array}}
\newcommand{\bqa}{\begin{eqnarray}}
\newcommand{\eqa}{\end{eqnarray}}
\newcommand{\bqaa}{\begin{eqnarray*}}
\newcommand{\eqaa}{\end{eqnarray*}}
\newcommand{\cO}{\mathcal{O}}
\newcommand{\bra}{\langle}
\newcommand{\ket}{\rangle}

\usepackage{color}

\begin{document}


\title {Radiative transition processes of light vector resonances in a chiral framework}

\author{ Yun-Hua~Chen$^a$}
\author{ Zhi-Hui~Guo$^b$}
\email[Corresponding author:]{ zhguo@mail.hebtu.edu.cn}
\author{ Han-Qing Zheng$^c$}
\affiliation{ ${}^a$  Institute of High Energy Physics, CAS, Beijing 100049, P.~R.~China
 \\ ${}^b$ Department of Physics, Hebei Normal University, Shijiazhuang 050024, P.~R.~China
 \\
 and State Key Laboratory of Theoretical Physics, Institute of Theoretical Physics, Chinese Academy of Sciences, Beijing 100190, P.R.China
 \\ ${}^c$ Department of Physics and State Key Laboratory of
Nuclear Physics and Technology,Peking University, Beijing 100871, P.~R.~China.  }

\begin{abstract}
Within the framework of chiral effective field theory,
we study various electromagnetic processes with light vector resonances: $K^{\ast}(892)\to K\gamma$, $e^{+}e^{-}\to K^{\ast}(892)K $ and the
$\omega\pi\gamma^\ast$ form factors. With two multiplets of vector resonances being introduced,
we fit the decay widths of $K^{\ast0}\to K^{0}\gamma$,
$K^{\ast+}\to K^{+}\gamma$, and the pertinent measurements from
the $e^{+}e^{-}\to K^{\ast \pm}(892)K^{\mp} $ cross sections, such as
moduli and relative phases between the isoscalar and isovector components from BABAR collaboration,
together with the $\omega\pi$ form factors from NA60, SND and CLEO collaborations.
The values of resonance couplings, masses and widths of the excited vector states $\rho'$ and $\phi'$ are then determined.
The $\omega'$-$\phi'$ mixing angle is discussed and turns out to be quite different from the ideal mixing case.
Three sources of $SU(3)$ symmetry breaking effects in the $\Gamma(K^{\ast}\to K\gamma)$ decays are identified and analyzed in detail.
\end{abstract}


\maketitle

\section{Introduction}

Electromagnetic transition form factor of light mesons is one of the key ingredients to study hadron properties and it recently gains intensive interests.
There are fruitful updated measurements from different experimental collaborations, such as NA60~\cite{NA6009,NA6011}, SND~\cite{SND2000,SND2013} and BABAR~\cite{babar08}.
On the theoretical side, the transition form factor provides us an important tool to study the intrinsic properties of light hadrons,
including light pseudoscalar mesons and vector resonances. Of more importance, it may also help to reduce the hadronic uncertainties in the light by light scattering, which
is an important source of theoretical uncertainties of the muon anomalous magnetic moment~\cite{Jegerlehner:2009ry,Bijnens:1995xf,Bijnens:2001cq}.

The transition form factor of $\pi^0\gamma^*\gamma^*$ is one of the most important form factors in the light by light scattering. Physics involved in this kind of
form factor is quite complicated, since one needs to handle different dynamics within a broad range of energy region. In the very high and low energy regions, we have
reliable and model-independent theoretical tools, namely pQCD and chiral perturbation theory ($\chi$PT).
While this is not the case in the intermediate energy region, where various resonances enter.
In the present work, we follow the chiral effective field theory explicitly including resonance states developed in Ref.~\cite{Ecker}
to study the radiative transition form factors involved with vector resonances.

$\chi$PT is a model-independent method to describe the QCD dynamics in the very low energy region ($E\ll M_\rho$), which is
based on chiral symmetry and expansions in terms of external
momentum and light quark masses~\cite{Weinberg}.
However the dynamical degrees of freedom in $\chi$PT are restricted to the light pseudo-Goldstone bosons $\pi, K, \eta$~\cite{gl845}.
In the intermediate energy region ($E\sim M_\rho$), clearly the resonance fields need to be explicitly included.
Refs.~\cite{Ecker,Ecker:1989yg} proposed an approach to incorporate resonances in a chiral covariant way.
In this theory, not only the chiral symmetry but also the QCD inspired high-energy behaviors at large $N_C$
are implemented, which makes the resonance chiral theory (R$\chi$T) bear more properties of QCD.
Moreover, to implement the high energy constraints in the chiral effective theory also makes it possible to apply the results from this theory directly
to some form factors with virtual particles, such as those in light by light scattering, where the QCD high energy behavior
can be important~\cite{Bijnens:1995xf}.
At the practical level, to impose the high energy constraints is an efficient way to reduce the number of free resonance coupling,
which makes R$\chi$T more predictive in the phenomenological discussions
~\cite{Cirigliano,Femenia,Jamins,Dumm,Mateu,Guo:2007ff,Guo,Guo:2010dv,chen,Roig:2013baa}.

In our previous work~\cite{chen}, we have performed an extensive study on the electromagnetic transition form factors and decays
of light pseudoscalar mesons $\pi,\eta,\eta'$ in the framework of R$\chi$T. In the present work, we focus on the similar form factors and decays
but involving light vector resonances. The relevant resonance operators in these kinds of processes are of the odd-intrinsic-parity type.
For the odd-intrinsic-parity sector, Ref.~\cite{Femenia} introduced a general effective chiral Lagrangian
containing symmetry allowed interactions between two vector objects (currents or lowest multiplet of resonances)
and one pseudoscalar meson, by employing the antisymmetric tensor formalism as
used in~Ref.~\cite{Ecker} to describe the vector resonances.
While in Ref.~\cite{knecht01epjc}, similar study was carried out
but the vector resonances was described in terms of the Proca vector field representation.
Later on, the vector-vector-pseudoscalar ($VVP$) type Lagrangian with vector resonances
in the antisymmetric tensor representation has been put forward in different
aspects: Ref.~\cite{Mateu} introduced a second nonet of vector resonances,
Ref.~\cite{kampf11prd} worked out the complete base of resonance operators that are
relevant to the $\cO{(p^6)}$ $\chi$PT Lagrangian in the anomaly sector
and in Ref.~\cite{chen} we have made a comprehensive discussion on the inclusion of the singlet $\eta_1$.
Though we see impressive progresses in this research field, one still needs to bear in mind that
in the strict large $N_C$ QCD there is an infinite tower of zero-width resonances. In practice,
typically one has to truncate the tower to the lowest multiplet of resonances, which is named as the minimal hadronical ansatz in Ref.~\cite{PerisdRafael}.
Under this approximation, R$\chi$T has been successfully applied to the phenomenological study on many processes where the
intermediate resonances play an important
role~\cite{Guo,Jamins,Dumm,chen,Guo:2010dv,Escribano:2013bca,Oller:2000ug,Escribano:2010wt,Guo:2011pa}.

However,  we notice that in the previous study of R$\chi$T, most efforts have been made on the lowest lying resonances. For
example, in the vector sector, the lowest nonet of $\rho(770)$,
$\omega(782)$, $K^*(892)$ and $\phi(1020)$ has been extensively studied, while investigation on the higher mass resonances is relatively rare.
So one of the important improvements of our present work is to study the excited vector resonances in chiral effective theory, comparing with
our previous work~\cite{chen}.
On the experimental side, recently the BABAR Collaboration has  measured the
$e^{+}e^{-}\rightarrow K^{\ast}(892)K $ cross sections from the threshold to the energy region around 2 $\sim$ 3~GeV~\cite{babar08},
which enables us to study properties of more massive vector resonances, i.e.,  $\rho'$, $\omega'$  and
$\phi'$. In this measurement, the moduli and relative phases of isoscalar and isovector components of
the $e^{+}e^{-}\rightarrow K^{\ast}(892)K $ cross sections are provided.
The updated data make a strong constraint on the free parameters in our theory and hence
allow us to accurately extract the resonance properties, such the masses and widths of the $\rho'$
and $\phi'$ (and also $\omega'$), their mixings, as well as their couplings to light mesons.
The $\omega'-\phi'$ mixing angle is estimated and we find that it is far from the ideal mixing case.

Another important issue we will address in this article is the $SU(3)$ symmetry breaking effect in radiative decays.
An ideal process to study this effect is $K^{\ast 0}\rightarrow K^{0}\gamma$ and
$K^{\ast\pm}\rightarrow K^{\pm}\gamma$.
It is well known that the ratio $\frac{\Gamma(K^{\ast 0}\rightarrow K^{0}\gamma)}{\Gamma(K^{\ast\pm}\rightarrow K^{\pm}\gamma)}=4$ in
the $SU(3)$ limit~\cite{Donnell}, while the world average value from experimental measurements is around $2.3$~\cite{Pdg}.
The large $SU(3)$ symmetry breaking effect has been discussed in many previous
works~\cite{Durso,Bagan,Bramon,Benayouna,Sakurai}. In the framework of R$\chi$T, it is interesting to point out
that there exists a special resonance operator
$O_{VJP}^4=\frac{ic_4}{M_V}\epsilon_{\mu\nu\rho\sigma}\bra V^{\mu\nu}[f_-^{\rho\sigma},\chi_+]\ket $~\cite{Femenia},
which contributes exclusively to the charged processes $K^{\ast\pm}\rightarrow K^{\pm}\gamma$.
Hence it provides an important source of $SU(3)$ symmetry breaking. In fact, Ref~\cite{Dumm} has determined
several values for the $c_4$ parameter in the discussion of hadronic $\tau$ decays.
Our present work provides another way to determine its value and allows us
to check which kinds of values from Ref.~\cite{Dumm} are reasonable.
A careful analysis of the strengths of different $SU(3)$ breaking terms $\Gamma(K^{\ast 0}\rightarrow K^{0}\gamma)$
and $\Gamma(K^{\ast\pm}\rightarrow K^{\pm}\gamma)$ will be delivered in our work.

Contrary to the $K^*\to K\gamma$ process, the $\omega\rightarrow \pi^0 \gamma^{*}$ form factor is free of
of large $SU(3)$ corrections. Nevertheless, another difficulty arises in this form factor, since it is found that
the well-established vector-meson-dominant (VMD) model fails to describe $\omega\rightarrow \pi^0 \gamma^*$~\cite{Klingl,Dzhelyadin,Faessler}.
In the present work, we discuss this transition form factor in R$\chi$T by including excited vector resonances in addition to the $\rho(770)$.
We will also confront our theoretical results with the new measurements from SND collaboration~\cite{SND2000,SND2013}.

This paper is organized as follows. In Sec.~\ref{theor}, we introduce the
theoretical framework and elaborate the calculations for $K^{\ast}(892)\rightarrow
K\gamma$ and $e^{+}e^{-}\rightarrow K^{\ast}(892)K $, the
$\omega\rightarrow \pi^0 \gamma^* $ transition form factor and the spectral
function for $\tau^- \rightarrow \omega\pi^- \nu_\tau$. In Sec.~\ref{disc}, we present the fit results and discuss the $SU(3)$ symmetry
breaking mechanism and the $\omega\pi$ form factors in detail. Summary and conclusions are given in Sec.~\ref{conclu}.

\section{Theoretical framework}
\label{theor}

\subsection{Resonance chiral theory and the odd intrinsic parity
effective lagrangian}\label{sect.lag}

The low-energy ($\Lambda <m_{\rho}\approx 700 $ MeV) dynamics of QCD
is ruled by the interaction of the octet of pseudoscalar mesons,
which are characterized by the spontaneous breaking of chiral symmetry. The
remarkably successful $\chi$PT~\cite{Weinberg}
describes the strong interactions among pseudoscalar mesons in a
model-independent way. The effective Lagrangian to lowest order,
$O(p^2)$, is given by
\begin{eqnarray}  \mathscr{L}_\chi^{(2)}=\frac{F^2}{4}\bra u_\mu u^\mu
+\chi_+\ket ,\end{eqnarray}
 where
\begin{eqnarray}  u_\mu =i[u^{+}(\partial _\mu \
-ir_\mu)u-u(\partial _\mu -il_\mu)u^{+}] , \nonumber
\end{eqnarray}
\begin{eqnarray} \chi_\pm
=u^{+}\chi u^{+}\pm u\chi^{+}u,
  \hspace{1.0cm} \chi=2B_0(s+ip) .
 \end{eqnarray}
The unitary matrix in flavour space
\begin{eqnarray}
u=\exp\{i\frac{\Phi}{\sqrt{2}F}\},
\end{eqnarray}
incorporates the pseudo-Goldstone octet:
\begin{equation}
\Phi=
 \left( {\begin{array}{*{3}c}
   {\frac{1}{\sqrt{2}}\pi ^0 +\frac{1}{\sqrt{6}}\eta _8 } & {\pi^+ } & {K^+ }  \\
   {\pi^- } & {-\frac{1}{\sqrt{2}}\pi ^0 +\frac{1}{\sqrt{6}}\eta _8} & {K^0 }  \\
   { K^-} & {\overline{K}^0 } & {-\frac{2}{\sqrt{6}}\eta_8 }  \\
\end{array}} \right).
\end{equation}
The external sources $r_\mu, l_\mu, s$ and $p$ promote the global
symmetry in the flavor space to a local one. The parameter $F$ corresponds to the pion decay constant $F_\pi=92.2$~MeV in the chiral limit and
$B_0$ is related to the quark condensate via $\bra 0\mid\psi\overline{\psi}\mid0\ket=-F^{2}B_{0}[1+O(m_{q})]$, being $m_q$ the light quark mass.
The explicit chiral symmetry breaking is implemented in $\chi$PT by taking the vacuum expectation values of the scalar sources as
$s=$Diag$\{m_u,m_d,m_s\}$. We work in the isospin limit throughout, i.e. taking $m_u=m_d$.

In a chiral covariant way, the lowest multiplet of vector meson resonances was explicitly included in Ref.~\cite{Ecker} in
terms of antisymmetric tensor fields. In this formalism, the kinetic term of the vector resonance Lagrangian takes the form
\begin{eqnarray}\label{lagkinv}
\mathscr{L}_{kin}(V)=-\frac{1}{2}\bra\nabla^{\lambda}V_{\lambda\mu}\nabla_{\nu}V^{\nu\mu}-\frac{M_V^2}{2}V_{\mu\nu}V^{\mu\nu}\ket\,,
\end{eqnarray}
with the ground vector nonet
\begin{equation}\label{defu3v}
V_{\mu\nu}=
 \left( {\begin{array}{*{3}c}
   {\frac{1}{\sqrt{2}}\rho ^0 +\frac{1}{\sqrt{6}}\omega _8+\frac{1}{\sqrt{3}}\omega _1 } & {\rho^+ } & {K^{\ast+} }  \\
   {\rho^- } & {-\frac{1}{\sqrt{2}}\rho ^0 +\frac{1}{\sqrt{6}}\omega _8+\frac{1}{\sqrt{3}}\omega _1} & {K^{\ast0} }  \\
   { K^{\ast-}} & {\overline{K}^{\ast0} } & {-\frac{2}{\sqrt{6}}\omega_8+\frac{1}{\sqrt{3}}\omega _1 }  \\
\end{array}} \right)_{\mu\nu}\,,
\end{equation}
and the covariant derivative and the chiral connection defined as
\begin{eqnarray}
\nabla_{\mu}V=\partial_{\mu}V+[\Gamma_{\mu},V],\hspace{1.5cm}
\Gamma_{\mu}=\frac{1}{2}\{u^{+}(\partial _\mu-ir_\mu)u+u(\partial
_\mu-il_\mu)u^{+}\}.
\end{eqnarray}
At leading order of $1/N_C$, the mass splitting of
the ground vector multiplet is governed by a single resonance operator~\cite{Cirigliano:2003yq}
\begin{eqnarray}
\label{lagemr}
-\,\,\,   \frac{1}{2} \, e_m^V \bra V_{\mu\nu} V^{\mu\nu} \chi_+ \ket \,.
\end{eqnarray}
In Ref.~\cite{guo09prd}, it has been demonstrated that this single operator can well 
explain the mass splittings of $\rho(770)$, $K^*(895)$  
and $\phi(1020)$. In addition to the mass splittings of these vector resonances, 
this operator is also shown to be able to perfectly reproduce the quark
mass dependences of the $\rho(770)$ mass from the Lattice simulation~\cite{Guo:2014yva}. 
We will then employ the single operator in Eq.~\eqref{lagemr} to account 
for the mass splittings of the lowest vector multiplet in this work. 
Additional $1/N_C$ suppressed operators, such as $\bra V_{\mu\nu}\ket \bra V^{\mu\nu} \ket$, 
will not be considered for the well-established ground vectors, due to the fact that $\omega(782)$ and $\phi(1020)$ 
in the ground multiplet are well described by the ideal mixing between the octet and singlet vector states. 
While because of the unclear situation for the excited vectors, 
the $1/N_C$ suppressed operators will be however introduced for them in later discussion. 
Combining Eqs.~\eqref{lagkinv} and \eqref{lagemr}, it can be easily verified that
the physical states of $\omega(782)$ and $\phi(1020)$ result from the ideal mixing of $\omega_1$ and $\omega_8$
\begin{eqnarray}\label{mixv}
\omega_1=\sqrt{\frac{2}{3}}\omega-\sqrt{\frac{1}{3}}\phi\,,
\hspace{2.5cm}
\omega_8=\sqrt{\frac{2}{3}}\phi+\sqrt{\frac{1}{3}}\omega\,.
\end{eqnarray}
Then the mass splitting pattern of the ground vector multiplet takes the form
\begin{eqnarray}\label{masssplitv}
M_{\rho}^2 = M_\omega^2&=&M_V^2-4e_m^V m_\pi^2\,, \nonumber \\
M_{K^*}^2&=&M_V^2-4e_m^V m_K^2\,, \nonumber \\
M_\phi^2&=&M_V^2-4e_m^V (2m_K^2-m_\pi^2)\,,
\end{eqnarray}
where we have used the leading order relations $2m_u B_0=m_\pi^2$ and $(m_u+m_s)B_0=m_K^2$.

For the general interaction Lagrangian linear in $V_{\mu\nu}$
up to $O(p^2)$, it reads~\cite{Ecker}
\begin{eqnarray}
\mathscr{L}_{2}(V)=\frac{F_V}{2\sqrt{2}}\bra V_{\mu\nu}f_{+}^{\mu\nu}\ket+\frac{iG_V}{\sqrt{2}}\bra V_{\mu\nu}u^\mu
u^\nu\ket,
\end{eqnarray}
\begin{eqnarray}
f_{\pm}^{\mu\nu}=uF_L^{\mu\nu}u^{+}\pm u^{+}F_R^{\mu\nu}u,
\end{eqnarray}
with $F_{L,R}^{\mu\nu}$ the field strength tensors of the left and
right external sources $l_\mu$ and $r_\mu$, respectively. $F_V, G_V$ are real resonance coupling constants and
only $F_V$ is relevant in our current study.

The interaction operators containing two-vector and one-pseudoscalar objects
have been worked out in Ref.~\cite{Femenia}
\begin{eqnarray}\label{lagvjp}
\mathscr{L}_{VJP}=&&\frac{c_1}{M_V}\epsilon_{\mu\nu\rho\sigma}\bra\{V^{\mu\nu},f_+^{\rho\alpha}\}\nabla_\alpha
u^\sigma\ket +\frac{c_2}{M_V}\epsilon_{\mu\nu\rho\sigma}\bra\{V^{\mu\alpha},f_+^{\rho\sigma}\}\nabla_\alpha
u^\nu\ket
+\frac{ic_3}{M_V}\epsilon_{\mu\nu\rho\sigma}\bra\{V^{\mu\nu},f_+^{\rho\sigma}\}\chi_-\ket  \nonumber\\&&
+\frac{ic_4}{M_V}\epsilon_{\mu\nu\rho\sigma}\bra V^{\mu\nu}[f_-^{\rho\sigma},\chi_+]\ket
 +\frac{c_5}{M_V}\epsilon_{\mu\nu\rho\sigma}\bra\{\nabla_\alpha
V^{\mu\nu},f_+^{\rho\alpha}\}u^\sigma\ket+\frac{c_6}{M_V}\epsilon_{\mu\nu\rho\sigma}\bra\{\nabla_\alpha
V^{\mu\alpha},f_+^{\rho\sigma}\}u^\nu\ket \nonumber\\&&
+\frac{c_7}{M_V}\epsilon_{\mu\nu\rho\sigma}\bra\{\nabla^\sigma
V^{\mu\nu},f_+^{\rho\alpha}\}u_\alpha\ket,
\end{eqnarray}
\begin{eqnarray}\label{lagvvp}
\mathscr{L}_{VVP}=&&d_{1}\epsilon_{\mu\nu\rho\sigma}\bra\{V^{\mu\nu},V^{\rho\alpha}\}\nabla_\alpha
u^\sigma\ket +id_{2}\epsilon_{\mu\nu\rho\sigma}\bra\{V^{\mu\nu},V^{\rho\sigma}\}\chi_-\ket\nonumber\\
&&+d_{3}\epsilon_{\mu\nu\rho\sigma}\bra\{\nabla_\alpha
V^{\mu\nu},V^{\rho\alpha}\}u^\sigma\ket +d_{4}\epsilon_{\mu\nu\rho\sigma}\bra\{\nabla^\sigma
V^{\mu\nu},V^{\rho\alpha}\}u_\alpha\ket\,,
\end{eqnarray}
with $\epsilon_{\mu\nu\rho\sigma}$ the Levi-Civita antisymmetric tensor.
Similar operators with an excited vector multiplet $V_1$ can be straightforwardly constructed ~\cite{Mateu}
\begin{eqnarray}\label{lagfv1}
\mathscr{L}_{2}(V_1)=\frac{F_{V_{1}}}{2\sqrt{2}}\bra V_{1\mu\nu}f_{+}^{\mu\nu}\ket,
\end{eqnarray}
\begin{eqnarray}\label{lagv1vp}
\mathscr{L}_{VV_{1}P}=&&d_{a}\epsilon_{\mu\nu\rho\sigma}\bra\{V^{\mu\nu},V_{1}^{\rho\alpha}\}\nabla_\alpha
u^\sigma\ket + d_{b}\epsilon_{\mu\nu\rho\sigma}\bra\{V^{\mu\alpha},V_{1}^{\rho\sigma}\}\nabla_\alpha
u^\nu \ket + d_{c}\epsilon_{\mu\nu\rho\sigma}\bra\{\nabla_\alpha
V^{\mu\nu},V_{1}^{\rho\alpha}\}u^\sigma\ket\nonumber\\ &&
+d_{d}\epsilon_{\mu\nu\rho\sigma}\bra\{\nabla_\alpha
V^{\mu\alpha},V_{1}^{\rho\sigma}\}u^\nu\ket + d_{e}\epsilon_{\mu\nu\rho\sigma}\bra\{\nabla^\sigma
V^{\mu\nu},V_{1}^{\rho\alpha}\}u_\alpha\ket+ id_{f}\epsilon_{\mu\nu\rho\sigma}\bra\{V^{\mu\nu},V_{1}^{\rho\sigma}\}\chi_-\ket\,. \nonumber \\
\end{eqnarray}
The kinetic Lagrangian for the excited vector takes the same form as that in Eq.~\eqref{lagkinv}.
The equivalent operators to Eq.~\eqref{lagvjp} involving the excited vector multiplet $V_1$ happen to be irrelevant to
the discussions in the present article.

For latter convenience, we follow the convention introduced in Ref.~\cite{Guo} to define certain combinations of $d_i$ in Eq.~\eqref{lagv1vp}
\begin{eqnarray}
d_m&=&d_a+d_b-d_c+8d_f,\nonumber\\
 d_M&=&d_b-d_a+d_c-2d_d,\nonumber\\
d_s&=&d_c+d_a-d_b.
\end{eqnarray}

In the present article, since we focus on the processes involving pion and kaon, the additional operators
that we introduced in Ref.~\cite{chen} to discuss $\eta$ and $\eta'$ will be irrelevant. So we do not elaborate more
on this issue. The flavor structure for the excited vector multiplet $V_1$ is the same as the ground multiplet in Eq.~\eqref{defu3v}.
While unlike the well-established entries in the ground vector multiplet, the contents of the excited vector multiplet are still not
clear. Therefore we include a general set of mass splitting operators to describe the excited vectors, instead of only using a single operator
in the ground multiplet case. Up to the linear quark mass corrections, the pertinent operators read~\cite{Cirigliano:2003yq}
\begin{eqnarray}\label{lagemv1}
\mathscr{L}_{V_1}^{mass-split}=-\frac{1}{2}\bigg( e_m^{V_1}\bra{V_1}_{\mu\nu}{V_1}^{\mu\nu}
\chi_+\ket -\frac{\gamma_{V_1} M_{V_1}^2}{2}\bra{V_1}_{\mu\nu}\ket \bra{V_1}^{\mu\nu}\ket\bigg)\,,
\end{eqnarray}
where the second operator is $1/N_C$ suppressed compared to the first one and $M_{V_1}$ is the mass of excited vector multiplet in chiral limit.
Another type operator, e.g. $-1/(2\sqrt{3})k_m^{V_1}\bra{V_1}_{\mu\nu}\ket\bra {\hat{V_1}}^{\mu\nu}\chi_+\ket$, was also
introduced in Ref.~\cite{Cirigliano:2003yq}, but we consider this operator is less relevant compared to the two terms in Eq.~\eqref{lagemv1}.
The reason is that compared to the $e_m^{V_1}$ term though the $\gamma_{V_1}$ term is $1/N_C$ suppressed, its chiral order is enhanced.
While for the $k_m^{V_1}$ term, its chiral and $1/N_C$ orders are both suppressed.

Owing to the $1/N_C$ suppressed operator in Eq.~\eqref{lagemv1}, the excited states $\omega'$ and $\phi'$ can not be simply described
as the ideal mixing of $\omega'_1$ and $\omega'_8$ as in the ground state case.
The physical states of $\omega'$ and $\phi'$ are related to the octet-singlet basis through
\begin{eqnarray}
\omega'=\sin\theta_V\omega'_8+\cos\theta_V\omega'_1\,, \qquad
\phi'=\cos\theta_V\omega'_8-\sin\theta_V\omega'_1\,.
\end{eqnarray}
The mixing angle $\theta_V$ and the masses of $\omega'$ and $\phi'$ can be expressed in terms of the couplings in Lagrangian~\eqref{lagemv1}
\begin{eqnarray}
M_{\omega^{'}}^2&=&\frac{M_{11}^2+M_{22}^2-\sqrt{(M_{11}-M_{22})^2+4M_{12}^2}}{2}\,, \\
M_{\phi^{'}}^2&=&\frac{M_{11}^2+M_{22}^2+\sqrt{(M_{11}-M_{22})^2+4M_{12}^2}}{2}\,,\\
\sin\theta_V&=&-\frac{M_{12}^2}{\sqrt{(M_{\phi^{'}}^2-M_{\omega^{'}}^2)(M_{11}^2-M_{\omega^{'}}^2)}}\,,
\end{eqnarray}
with
\begin{eqnarray}
M_{11}^2&=& {M_{V_1}^2-\frac{4}{3}e_m^{V_1}(4m_K^2-m_\pi^2) } \,,  \\
M_{22}^2&=&{M_{V_1}^2(1+\gamma_{V_1})-\frac{4}{3}e_m^{V_1}(2m_K^2+m_\pi^2)  } \,, \\
M_{12}^2&=& {\frac{8\sqrt{2}}{3}e_m^{V_1}(m_K^2-m_\pi^2) } \,.
\end{eqnarray}
While for the masses of $\rho'$ and $K^{*'}$, they are only contributed by the $e_m^{V_1}$ term in Eq.~\eqref{lagemv1}
and take the same form as the ground states in Eq.~\eqref{masssplitv}
\begin{eqnarray}
M_{\rho'}^2=M_{V_1}^2-4e_m^{V_1}m_\pi^2\,,\\
M_{K^{*'}}^2=M_{V_1}^2-4e_m^{V_1}m_K^2\,.
\end{eqnarray}

To include the quark mass corrections to the resonance masses in the Lagrangian approach allows us
to incorporate the $SU(3)$ symmetry breaking effects in a systematic way. This is important to understand
the $SU(3)$ symmetry breaking mechanism in the $K^*K$ transitions.
Not only the quark mass corrections to the resonance masses but also to interacting vertexes are important to
fully study the $SU(3)$ symmetry breaking effects.
The $VVP$-type Lagrangians introduced in Eqs.~\eqref{lagvjp}\eqref{lagvvp}\eqref{lagv1vp}, which are constructed
by using the chiral building bricks of the pseudo-Goldstone mesons, provides us an appropriate framework to systematically take into account the relevant
interacting vertexes. Though it consists of a large number of free couplings, we will see later that in a given process not
all of them are independent. In fact in our present discussion, at most only six of the combinations of the $VVP$-type
couplings will appear after taking into account the high energy constraints dictated by QCD.

To impose the high energy constraints will not only make the R$\chi$T bear more properties
from QCD, but will also be helpful to reduce the number of free parameters. To proceed, we match the leading operator product
expansion (OPE) of the $VVP$ Green function with the result evaluated within R$\chi$T and require the vector form factor
vanish in the high energy limit. This leads to the following constraints on resonance couplings
\begin{eqnarray}
4c_3+c_1&=&0,\nonumber\\
c_1-c_2+c_5&=&0,\nonumber\\
 c_6-c_5&=&\frac{2d_3F_V+d_sF_{V_1}}{2\sqrt{2}M_V}\,,
\label{eqconstraint1}
\end{eqnarray}
which are already given by one of us in Ref.~\cite{Guo}. We point out that the previous constraints are obtained in the chiral and large $N_C$ limits, as
done in most of the R$\chi$T study~\cite{Cirigliano,Femenia,Jamins,Dumm,Mateu,Guo:2007ff,Guo,Guo:2010dv,chen,Roig:2013baa} in the literature.

\subsection{The transition amplitudes of $K^{\ast}\longrightarrow K \gamma $ decays and
$e^{+}e^{-}\longrightarrow K^{\ast} K $ processes}

\begin{figure*}[ht]
\begin{center}
\includegraphics[width=0.8\textwidth]{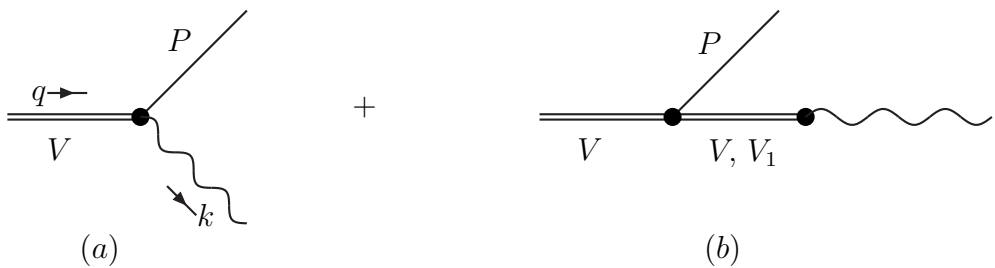}
\caption{ Diagrams relevant to the $V\rightarrow P \gamma(\gamma^{*})$
processes: (a) direct type and (b) indirect type. } \label{figVPgamma}
\end{center}
\end{figure*}

At leading order of $1/N_C$ in R$\chi$T, the $V\longrightarrow P \gamma(\gamma^{*})$ transitions receive two
types of contributions as displayed in Fig.~\ref{figVPgamma}, namely the direct type depicted by the diagram (a)
 and the indirect one by the diagram (b).

The structure of the amplitude for the~$K^{\ast}(q)\rightarrow K + \gamma(k)$~process can be written as
\begin{eqnarray} \label{defkskgamp}
i\mathscr{M}_{K^{\ast}K \gamma} &\equiv&
i\epsilon_{\mu\nu\rho\sigma}\epsilon_{K^{\ast}}^\mu
\epsilon_\gamma^\nu q^\rho k^\sigma e  g_{K^{\ast}K \gamma}^{total}\,,
\nonumber \\
g_{K^{\ast}K \gamma}^{total}&=& g_{K^{\ast}K \gamma}^{direct} + g_{K^{\ast}K \gamma}^{indirect}\,,
\end{eqnarray}
where $q$ and $k$ denote the four momenta of $K^{\ast}$ and $\gamma$ respectively,
$\epsilon_{K^{\ast}}$ and $\epsilon_{\gamma}$ correspond to the polarization vectors in that order;
$e$ stands for the electric charge of a positron. In the previous equation, we collect the contributions from the direct
and indirect  diagrams of Fig.~\ref{figVPgamma} in the quantities $g_{K^{\ast}K \gamma}^{direct}$ and
$g_{K^{\ast}K \gamma}^{indirect}$ respectively.

By using the previously introduced resonance Lagrangian in Sec.~\ref{sect.lag}, it is straightforward to
calculate $g_{K^{\ast}K \gamma}^{direct}$ and $g_{K^{\ast}K \gamma}^{indirect}$:
\begin{eqnarray}
g_{K^{\ast0}K^{0}\gamma}^{direct}&=& -\frac{4\sqrt{2}}{3F_K M_V
M_{K^*}}  [(c_1+c_2+8c_3-c_5)m_K^2 +(c_2+c_5 -c_1-2c_6)M_{K^*}^2]\,,
\nonumber\\
g_{K^{\ast0}K^{0}\gamma}^{indirect}&=&-\frac{2 F_V}{3F_K M_{K^*}} (
\frac{1}{M_\omega^2} -\frac{3}{M_\rho^2}-\frac{2}{M_\phi^2} )[(
d_1+8d_2-d_3)m_K^2  +d_3 M_{K^*}^2] \,, \nonumber\\&&
-\frac{F_{V_1}}{3F_K M_{K^*}} \bigg[
\frac{(2\sqrt{2}\cos\theta_V-\sin\theta_V)\sin\theta_V}{M_{\omega'}^2}
-\frac{3}{M_{\rho'}^2}
-\frac{(\cos\theta_V+2\sqrt{2}\sin\theta_V)\cos\theta_V}{M_{\phi'}^2}
\bigg] \nonumber\\&& (d_m m_K^2+d_M M_{K^*}^2)\,,
\label{eqKstar0K}
\end{eqnarray}
\begin{eqnarray}
g_{K^{\ast+}K^{+}\gamma}^{direct}&=& \frac{2\sqrt{2}}{3F_K M_V
M_{K^*}}  [(c_1+c_2+8c_3-c_5)m_K^2 +(c_2+c_5
-c_1-2c_6)M_{K^*}^2\nonumber\\
&&+24c_4(m_K^2-m_\pi^2)] \,,
\nonumber\\
g_{K^{\ast+}K^{+}\gamma}^{indirect}&=&-\frac{2 F_V}{3F_K M_{K^*}} (
\frac{1}{M_\omega^2} +\frac{3}{M_\rho^2}-\frac{2}{M_\phi^2} )[(
d_1+8d_2-d_3)m_K^2  +d_3 M_{K^*}^2]\nonumber\\&&
-\frac{F_{V_1}}{3F_K M_{K^*}} \bigg[
\frac{(2\sqrt{2}\cos\theta_V-\sin\theta_V)\sin\theta_V}{M_{\omega'}^2}
+\frac{3}{M_{\rho'}^2}
-\frac{(\cos\theta_V+2\sqrt{2}\sin\theta_V)\cos\theta_V}{M_{\phi'}^2}
\bigg]\nonumber\\&& (d_m m_K^2+d_M M_{K^*}^2)\,.
\label{eqKstarpK}
\end{eqnarray}
Physical meson masses are used in the kinematics. The kaon
decay constant $F_K$ has been employed in the amplitude, instead of the pseudo-Goldstone decay constant $F$ in chiral limit that
appears in the resonance chiral Lagrangian. We will take $F_K=0.113 $ GeV from PDG~\cite{Pdg}.

Within our formalism, the $\Gamma(K^{\ast}\rightarrow K\gamma)$ decay width is given by
\begin{eqnarray}
\Gamma(K^{\ast}\rightarrow
K\gamma)=\frac{1}{3}\alpha\bigg(\frac{M_{K^{\ast}}^2-m_K^2}{2M_{K^{\ast}}}\bigg)^3{\mid
g_{K^{\ast}K \gamma}^{total}\mid}^2\,,
\end{eqnarray}
with $\alpha=e^2/(4\pi)$ the fine structure constant.

\begin{figure*}[ht]
\begin{center}
\includegraphics[width=1.0\textwidth]{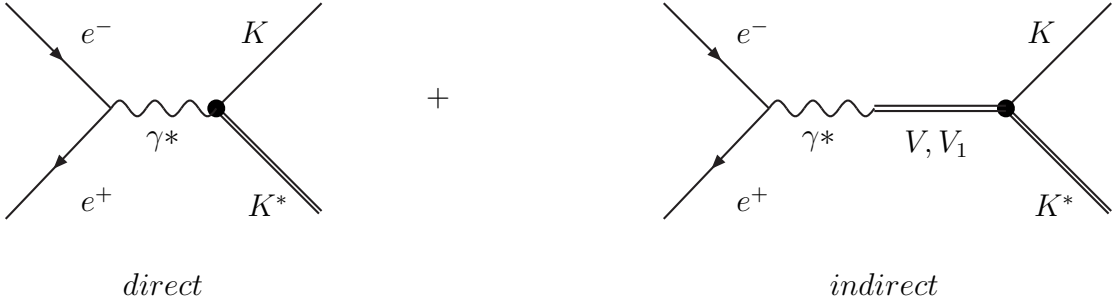}
\caption{ Relevant diagrams to the process $e^{+}e^{-}\longrightarrow
K^{\ast}(892)K. $ } \label{figeeVP}
\end{center}
\end{figure*}

In the leading order of large $N_C$, the $e^{+}e^{-}\rightarrow K^{\ast}(892)K $
transitions also receives two types of contributions, i.e. direct and indirect types, as displayed in Fig.~\ref{figeeVP}

The main difference between $e^{+}e^{-}\rightarrow K^{\ast}(892)K $ and $K^{\ast}\rightarrow K\gamma$ is that
now we have an off-shell photon. The final results for the amplitudes of $ \gamma^{\ast}\longrightarrow K^{\ast }K$
can be obtained in the same way as the $K^{\ast}\rightarrow K\gamma$ case.
In order to confront our theoretical results with the experimental data, we need to work in the isospin bases for
the amplitudes
\begin{eqnarray}\label{ampi0}
i M_{I=0}=&&i \frac{\sqrt{2}}{2}(M_{\gamma^{\ast}\longrightarrow
K^{\ast +}K^-}+M_{\gamma^{\ast}\longrightarrow \overline{K}^{\ast
0}K^0}) \nonumber\\ =&&
i\epsilon_{\alpha\beta\rho\sigma}\epsilon_{K^\ast}^{\alpha}\epsilon_{\gamma}^{\beta}q^{\rho}k^{\sigma}e\frac{\sqrt{2}}{2}
\frac{1}{F_{K}M_{K^{\ast}}}\bigg\{-\frac{2\sqrt{2}}{3M_{V}}[(c_1+c_2+8c_3-c_5)M_{K}^2+(c_2+c_5-c_1-2c_6)M_{K^{\ast}}^2\nonumber \\
&&+(c_1-c_2+c_5)s -24c_4(m_K^2-m_{\pi}^2)]-4F_{V}\bigg[
\frac{1}{3(M_{\omega}^2-s -iM_{\omega}\Gamma_{\omega})}
-\frac{2}{3(M_{\phi}^2-s-iM_{\phi}\Gamma_{\phi})} \bigg] \nonumber
\\&&
[(d_1+8d_2)m_K^2+d_3(M_{K^{\ast}}^2+s-m_K^2)] -2F_{V_1} \bigg[
\frac{(2\sqrt{2}\cos\theta_V-\sin\theta_V)\sin\theta_V}
{3(M_{\omega'}^2-s-iM_{\omega'}\Gamma_{\omega'}(s))}
\nonumber \\
&&-\frac{(\cos\theta_V+2\sqrt{2}\sin\theta_V)\cos\theta_V}
{3(M_{\phi'}^2-s-iM_{\phi'}\Gamma_{\phi'}(s))} \bigg]
 (d_m m_K^2+d_M M_{K^{\ast}}^2+d_s s)
\bigg\}\nonumber \\
\triangleq&&i\epsilon_{\alpha\beta\rho\sigma}\epsilon_{K^\ast}^{\alpha}\epsilon_{\gamma}^{\beta}q^{\rho}k^{\sigma}eA_0,
\end{eqnarray}
\begin{eqnarray}\label{ampi1}
i M_{I=1}=&&i \frac{\sqrt{2}}{2}(M_{\gamma^{\ast}\longrightarrow
\overline{K}^{\ast 0}K^0}-M_{\gamma^{\ast}\longrightarrow K^{\ast
+}K^-}) \nonumber\\
=&&i\epsilon_{\alpha\beta\rho\sigma}\epsilon_{K^\ast}^{\alpha}\epsilon_{\gamma}^{\beta}q^{\rho}k^{\sigma}e\frac{\sqrt{2}}{2}
\frac{1}{F_{K}M_{K^{\ast}}} \bigg\{-\frac{2\sqrt{2}}{M_{V}}[(c_1+c_2+8c_3-c_5)M_K^2+(c_2+c_5-c_1-2c_6)M_{K^{\ast}}^2\nonumber \\
&&+(c_1-c_2+c_5)s +8c_4(m_K^2-m_{\pi}^2)]+4F_{V}
\frac{1}{M_{\rho}^2-s-iM_{\rho}\Gamma_{\rho}(s)}
[(d_1+8d_2)m_K^2\nonumber \\
&&+d_3(M_{K^{\ast}}^2+s-m_K^2)]
 +2F_{V_1}
\frac{1}{M_{\rho'}^2-s-iM_{\rho'}\Gamma_{\rho'}(s)} (d_m
m_K^2+d_M M_{K^{\ast}}^2+d_s s)
\bigg\}\nonumber \\
\triangleq&&i\epsilon_{\alpha\beta\rho\sigma}\epsilon_{K^\ast}^{\alpha}\epsilon_{\gamma}^{\beta}q^{\rho}k^{\sigma}eA_1\,,
\end{eqnarray}
where $q$ and $k$ stand for the momenta of $K^{\ast}$ and $\gamma^\ast$ respectively and $s=k^2$ is the energy square of
the $e^+e^-$ system in center of mass frame.
The energy-dependent decay widths for intermediate resonances, such as $\Gamma_{\rho}(s),\Gamma_{\rho'}(s),\Gamma_{\phi'}(s)$ and $\Gamma_{\omega'}(s)$,
can be important, since the widths of
the relevant resonances are typically not so narrow and the final results can be sensitive to the off-shell width effects.
To rigorously include the off-shell width effects one needs to carry out the next-to-leading-order of $1/N_C$ computation
in R$\chi$T, which is beyond the scope of this work, since we focus on the calculations of the various amplitudes at leading order of $1/N_C$.
Nevertheless, in order to quantitatively estimate to what extent the off-shell width effects will
affect the final outputs, we will study two different forms of the energy-dependent widths, which will be discussed in detail in next section.

We see that only the $\omega-$like and $\phi-$like intermediate mesons contribute to the $M_{I=0}$ and only
the $\rho-$like intermediate mesons contribute to the $M_{I=1}$. While for the local terms or the so-called background terms in many other works~\cite{babar08}, they
are obtained from a Lagrangian based theory in this work, not simply introduced by hand.
The cross sections are related to the isospin amplitudes through
\begin{eqnarray} \sigma_{e^+e^-\longrightarrow K^{\ast
}K(I=0,1)} =\frac{\pi\alpha^2|A_{I=0,1}|^2
}{6s^3}(s^2+M_{K^{\ast}}^4+m_K^4-2sM_{K^{\ast}}^2-2sm_K^2-2M_{K^{\ast}}^2m_K^2)^{\frac{3}{2}},
\end{eqnarray}
where the masses of electron and positron have been neglected.

\subsection{The $\omega\longrightarrow \pi^0 \gamma^* $ transition form factor and the spectral function for $\tau^- \rightarrow \omega\pi^-
\nu_\tau$}

The amplitude for the radiative decay $\omega(q) \to \pi^0(p)
\gamma^{*}(k)$ can be written as:
\begin{eqnarray}\label{defFF}
i\mathcal{M}_{\omega\to \pi^0\gamma^{*}}= i \,e\,
\varepsilon_{\mu\nu\rho\sigma} \epsilon_\omega^\mu
\epsilon_{\gamma^{*}}^\nu q^\rho k^\sigma f_{\omega\pi^0}(s)\,,
\end{eqnarray}
where $s=k^2$, $\epsilon_\omega$ and $\epsilon_{\gamma^*}$ denote
the polarization vectors of the $\omega$ resonance and the off-shell
photon respectively.
The explicit expression of the electromagnetic transition form factor $f_{\omega\pi^0}(s)$ is
\begin{eqnarray}\label{omegaff}
f_{\omega\pi^0}(s)&=&-\frac{2\sqrt2 }{F_\pi M_V M_\omega} \big[
(c_1+c_2+8c_3-c_5)m_\pi^2 +(c_2+c_5 -c_1-2c_6)M_\omega^2
+(c_1-c_2+c_5)s \,\big] \nonumber
\\&& +\frac{4 F_V}{F_\pi M_\omega }\, \frac{1}{M_{\rho}^2-s-iM_{\rho}\Gamma_{\rho}(s)} \, \big[
(d_1+8d_2-d_3)m_\pi^2 +d_3 (M_\omega^2 + s) \big] \nonumber
\\&& +\frac{2
F_{V_1}}{F_\pi M_\omega }\,
\frac{1}{M_{\rho'}^2-s-iM_{\rho'}\Gamma_{\rho'}(s)} \,
\big[ d_M M_\omega^2 +d_s s +d_m m_\pi^2 \big]\,.
\end{eqnarray}

In the phenomenological discussion, it is common to normalize the form factor $f_{\omega\pi^0}(s)$
by its value at $s=0$. So the interesting quantity that we will study later is
\begin{eqnarray}\label{defFFnl}
F_{\omega\pi^0}(s)=\frac{f_{\omega\pi^0}(s)}{f_{\omega\pi^0}(0)}.
\end{eqnarray}

The charged $\omega\pi$ form factor or the $\omega\pi$ spectral function can be
measured in the semileptonic decays of the $\tau$ lepton. In the Eq.(35) of Ref.~\cite{Guo},
one of us has calculated the $\tau^- \rightarrow \omega\pi^- \nu_\tau$ spectral function.
We give the explicit expression here for completeness
\begin{eqnarray}
V(s)&=&\frac{1}{6F^2 M_\omega^2 \pi s^2 S_{EW}}\times
[m_\pi^4+(M_\omega^2-s)^2-2m_\pi^2(M_\omega^2-s)]^{3/2}\nonumber\\&&
\times \bigg|(2d_3 F_V+d_s F_{V_1})\frac{M_\omega^2}{M_V^2}+F_{V_1}(d_m
m_\pi^2+d_M M_\omega^2 +d_s
s)\frac{1}{M_{\rho'}^2-s-iM_{\rho'}\Gamma_{\rho'}(s)}\nonumber\\&&+2F_V[(d_1+8d_2)m_\pi^2
+d_3
(s+M_\omega^2-m_\pi^2)]\frac{1}{M_{\rho}^2-s-iM_{\rho}\Gamma_{\rho}(s)}\bigg|^2,
\end{eqnarray}
where the electroweak correction factor $S_{EW}$ has been analyzed
in~\cite{SEW}, and its value will be taken as $S_{EW}=1.0194$.

\section{Phenomenological discussions }
\label{disc}

\subsection{The fit results}

The experimental data that we consider here consist of eight different types:
$\Gamma(K^{\ast0}\rightarrow K^{0}\gamma)$~\cite{Pdg}, $\Gamma(K^{\ast+}\rightarrow K^{+}\gamma)$~\cite{Pdg},
the $e^{+}e^{-}\rightarrow K^{\ast \pm}(892)K^{\mp} $ cross sections~\cite{babar08}, the
moduli and relative phases of isoscalar and isovector components
of the $e^{+}e^{-}\rightarrow K^{\ast}(892)K $ cross sections~\cite{babar08}, the $\omega \rightarrow\pi^0 \gamma^*$
transition form factor~\cite{LG,NA6009,NA6011,SND2000,SND2013} and
the $\tau^- \rightarrow \omega\pi^- \nu_\tau$ spectral
function~\cite{CLEO}.

For the $e^{+}e^{-}\rightarrow K^{\ast}(892)K$ data in Ref.~\cite{babar08},
since in the energy region above 1.8 GeV the contributions coming from
other excited vector resonances with higher mass and spin, certainly affect
the $KK^{\ast}(892)$ channel, we only fit the data below 1.8 GeV.
For the $\omega'$ [$\omega(1420)$], we fix its mass and width at their world average
values, $M_{\omega'}$=1.42 GeV and $\Gamma_{\omega'}=0.215$GeV~\cite{Pdg},
because its mass is quite close to the $K^\ast K$ threshold
and its contribution only marginally shows up in the left tale of the $e^{+}e^{-}\rightarrow K^{\ast}(892)K$ cross sections.

Due to the fact that the masses of the ground vector resonances, $\rho,\omega,\phi,K^*$,
can be perfectly described by Eq.~\eqref{masssplitv}, see Refs.~\cite{guo09prd,Guo:2014yva}, we simply employ the physical masses of $\rho,\omega,\phi,K^*$
in the numerical discussions, instead of fitting the $e_m^V$ coupling. We point out that if the value of $e_m^V$ is fitted, it turns out to be very close
to the values given in Refs.~\cite{guo09prd,Guo:2014yva}.
For the mass of the ground vector in chiral limit, we take the value $M_V=764.3$~MeV from Ref.~\cite{guo09prd}.
About the value of $F_V$, it has been accurately determined by taking into account
a large amount of experimental data in our previous work~\cite{chen}. Hence we will take the value from that reference in this work, which is $F_V=0.137$~GeV.
For $F_{V_1}$ in Eq.~\eqref{lagfv1}, since what appears in our discussion is always the combination of
$F_{V_1}(d_m m_K^2+d_M M_{K^{\ast}}^2)$ or $F_{V_1}(d_m m_K^2+d_M M_{K^{\ast}}^2+d_s s)$,
which is already noticed in Ref.~\cite{Guo},  we can fix the sign of $F_{V_1}$ and leave $d_m, d_M$ and $d_s$ for free.
In Ref.~\cite{Guo} $F_{V_1}=-0.1$ GeV has been determined and we will take this value in our current study.

As mentioned previously, we introduce the finite-width effects to the intermediate resonances.
To systematically include the finite-width contributions in R$\chi$T, one has to step into the next-to-leading order of $1/N_C$ computation, e.g the one-loop calculations,
which is far beyond the scope of our study. Nevertheless, we will employ two different parameterizations for the energy-dependent widths for the broad resonances
in order to see how important the off-shell width effects will affect the fit results.
For the narrow-width resonances, such as $\omega$ and $\phi$, we simply use the constant widths in their propagators.

For the first case, which will be referred as Fit I in later discussion, we construct the energy-dependent widths for
broad resonances following the approach given in Ref.~\cite{GomezDumm:2000fz}, where the authors have taken
the chiral symmetry, analyticity and unitarity into account when discussing the form
of the off-shell width for the $\rho(770)$ resonance. The result of the previous reference has been generalized to other
resonances in many phenomenological discussions~\cite{Guo,chen,Jamins,Arganda}.
In this formalism, we construct the energy-dependent widths in the following way:
\begin{eqnarray} \label{widthi}
\Gamma_{\rho}(s)&=&\frac{sM_\rho}{96\pi
F_\pi^2}[\sigma_{\pi\pi}^3(s) +\frac{1}{2}\sigma_{KK}^3(s) ]\,, \nonumber \\
\Gamma_{\rho'}(s)&=&\Gamma_{\rho'}\frac{s}{M_{\rho'}^2}[\frac{\sigma_{\pi\pi}^3(s)+\frac{1}{2}\sigma_{KK}^3(s)}
{\sigma_{\pi\pi}^3(M_{\rho'}^2)+\frac{1}{2}\sigma_{KK}^3({M_{\rho'}}^2)}]\,, \nonumber \\
\Gamma_{\omega'}(s)&=&\Gamma_{\omega'}\frac{s}{M_{\omega'}^2}[\frac{\sigma_{\rho\pi}^3(s)
+\frac{(2\sqrt{2}\cos\theta_V-\sin\theta_V)^2}{3(\sqrt{2}\cos\theta_V+\sin\theta_V)^2}\sigma_{K^{\ast}K}^3(s)}
{\sigma_{\rho\pi}^3({M_{{\omega'}}^2})
+\frac{(2\sqrt{2}\cos\theta_V-\sin\theta_V)^2}{3(\sqrt{2}\cos\theta_V+\sin\theta_V)^2}
\sigma_{K^{\ast}K}^3(M_{{\omega'}}^2)}]\,, \nonumber \\
\Gamma_{\phi'}(s)&=&\Gamma_{\phi'}\frac{s}{M_{\phi'}^2}[\frac{\sigma_{K^{\ast}K}^3(s)
+\frac{4(\sqrt{2}\cos\theta_V+\sin\theta_V)^2}{3(\cos\theta_V+2\sqrt{2}\sin\theta_V)^2}\sigma_{\phi\eta}^3(s)}
{\sigma_{K^{\ast}K}^3({M_{{\phi'}}^2})
+\frac{4(\sqrt{2}\cos\theta_V+\sin\theta_V)^2}{3(\cos\theta_V+2\sqrt{2}\sin\theta_V)^2}
\sigma_{\phi\eta}^3(M_{{\phi'}}^2)}]\,,
\end{eqnarray}
where
\begin{eqnarray}
\sigma_{QR}(s)=\frac{1}{s}\sqrt{(s-(m_Q+m_R)^2)(s-(m_Q-m_R)^2)}\times
\theta(s-(m_Q+m_R)^2)\,,
\end{eqnarray}
with $\theta(s)$ the standard step function.

In Fit II, we take a different asymptotic behavior for the energy-dependent widths as used in Fit I
\begin{eqnarray}
 \Gamma_{R}(s) = \Gamma_{R}^{\rm I}(s)\frac{M_R}{\sqrt{s}}\,,  \quad  R=\rho\,,\rho'\,,\omega'\,,\phi'\,,
\end{eqnarray}
with $\Gamma_{R}^{\rm I}(s)$ defined in Eq.~\eqref{widthi}.
In the really asymptotic region when $s\to \infty$, the power of the energy-dependent width should be less than one, in such
a way that the phase of the amplitudes in Eqs.~\eqref{ampi0} and \eqref{ampi1} would approach to $\pi$. In Fit I, we adopt the results from Ref.~\cite{GomezDumm:2000fz},
where only $\pi\pi$ and $\bar{K}K$ channels are considered when deriving the finite width for $\rho(770)$.
In principle infinite number of channels will be open when $s\to \infty$, which could finally alternate the asymptotic behaviors obtained by considering finite number of channels.
A definite answer to this problem clearly deserves an independent work. As a phenomenological study, we will exploit both forms to perform the fits and examine to what extent different
forms can affect the final outputs.

\begin{table}[ht]
\begin{scriptsize}
\begin{center}
\begin{tabular}{|c||c|c|}
\hline
         & Fit I & Fit II   \\
\hline \hline
$M_{V_1} $     &   $ 1587\pm 7 $  & $1523\pm 5 $ \\
$e_m^{V_1}$     &   $ -0.262\pm 0.014 $     &     $-0.329\pm 0.010 $\\
$\gamma_{V_1}$   &       $ -0.246\pm 0.006 $     &     $-0.180\pm 0.006 $\\
$\Gamma_{\rho^{'}}$  (MeV)   &      $ 545\pm 22 $    &     $433\pm 13 $\\
$\Gamma_{\phi^{'}}$  (MeV)   &      $ 266\pm 17 $    &     $210\pm 11 $\\
$c_4$       &        $-0.0023\pm 0.0006$ &$-0.0024\pm 0.0005$ \\
$d_3$         &        $-0.198\pm 0.004$ &$-0.191\pm 0.004$ \\
$d_M$       &        $-0.38\pm 0.10$     &$-0.21\pm 0.08$ \\
$d_s$         &     $ -0.13\pm 0.02  $   &$-0.12\pm 0.02$ \\
$d_m$         &     $ -1.79\pm 0.15  $   &$-1.22\pm 0.12$ \\
$d_1+8d_2$   &        $-0.38\pm 0.04   $  &     $-0.36\pm 0.04$ \\
\hline
$\frac{\chi^2}{d.o.f}$   & $\frac{467.7}{210-11}=2.35$ & $\frac{589.1}{210-11}=2.96$ \\
\hline
Results for $V_1$ &&\\
\hline
$M_{\rho'} $ (MeV)    &   $ 1593\pm 8 $  & $1531\pm 8 $ \\
$M_{\phi'}$ (MeV)    &        $ 1709\pm 7 $     &     $1690\pm 6 $\\
$M_{{K^{\ast}}'}$ (MeV)    &        $ 1667\pm 6 $     &     $1626\pm 7 $\\
$\theta_{V}$ &  $15.1^{\circ}\pm 2.0^{\circ}  $ & $21.3^{\circ}\pm 2.2^{\circ} $\\
\hline
\end{tabular}
\caption{\label{tablepar} The parameters result from Fit~I and Fit~II.}
\end{center}
\end{scriptsize}
\end{table}

\begin{table}[ht]
\begin{footnotesize}
\begin{center}
\begin{tabular}{ccccc}\hline\hline
&Exp. &         Fit I  & Fit II      \\
\hline
$\Gamma_{K^{*0}\rightarrow K^0\gamma}$($\times 10^{-5}$ GeV) & $11.6\pm1.2$      & $13.1\pm1.3$&$14.1\pm 1.2$\\
$\Gamma_{K^{*+}\rightarrow K^+\gamma}$($\times 10^{-5} $GeV) & $5.0\pm0.6$      & $5.0\pm 1.0$&$5.3\pm 1.0$\\
$\frac{\Gamma_{K^{*0}\rightarrow K^0\gamma}}{\Gamma_{K^{*+}\rightarrow K^+\gamma}}$ & $2.6\pm0.3$       & $2.6\pm 0.4$&$2.7\pm 0.5$\\
\hline\hline
\end{tabular}
\caption{ \label{tableKstarK} Experimental and fit values of
$\Gamma(K^{\ast0}\rightarrow K^{0}\gamma)$,
$\Gamma(K^{\ast+}\rightarrow K^{+}\gamma)$ and
$\frac{\Gamma_{K^{*0}\rightarrow
K^0\gamma}}{\Gamma_{K^{*+}\rightarrow K^+\gamma}}$. }
\end{center}
\end{footnotesize}
\end{table}

\begin{figure*}[ht]
\begin{center}
\includegraphics[width=1.0\textwidth]{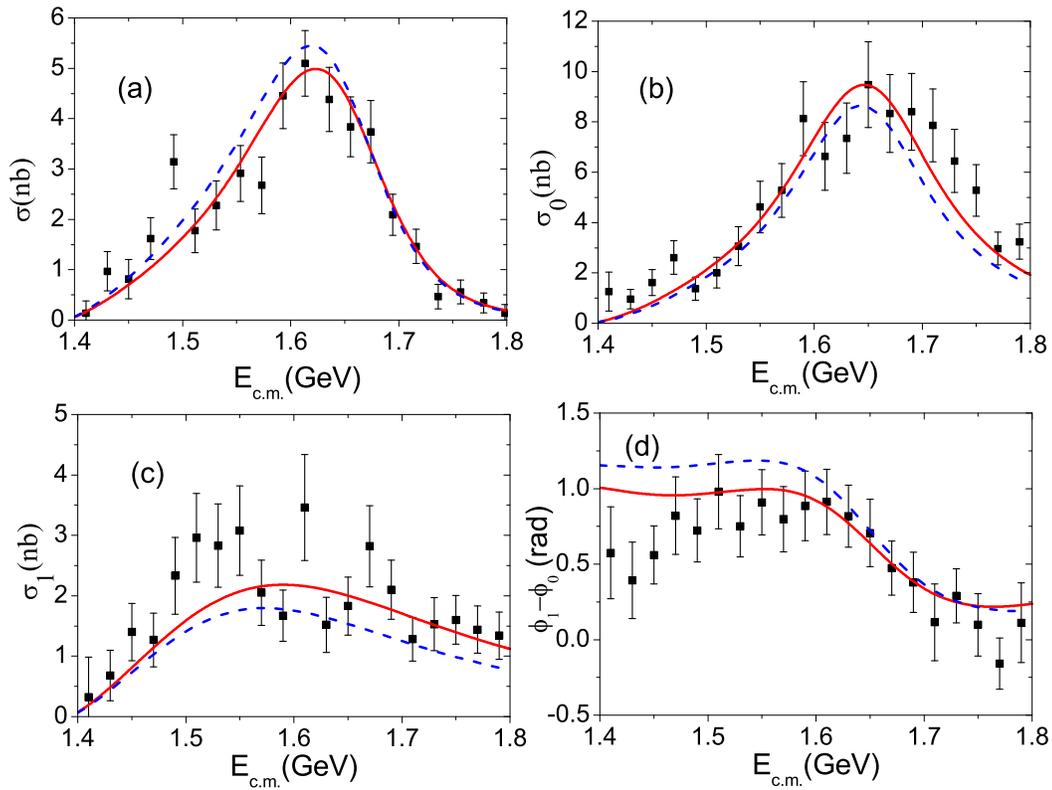}
\caption{ The fit results of Fit~I~( red solid line) and Fit~II~(blue dashed line). Panel (a) is to display the cross
sections of $\sigma_{K^{\ast \pm}K^{\mp}}(892)$. Panel (b) is to display the isoscalar
components of the $\sigma_{K^{\ast}K}$. Panel (c) is to display isovector components of
the $\sigma_{K^{\ast}K}$. Panel (d) is to display the phase differences between isovector and
isoscalar $K^{\ast}(892)K$ amplitudes.   }\label{figkstark}
\end{center}
\end{figure*}

\begin{figure*}[ht]
\begin{center}
\includegraphics[width=1.0\textwidth]{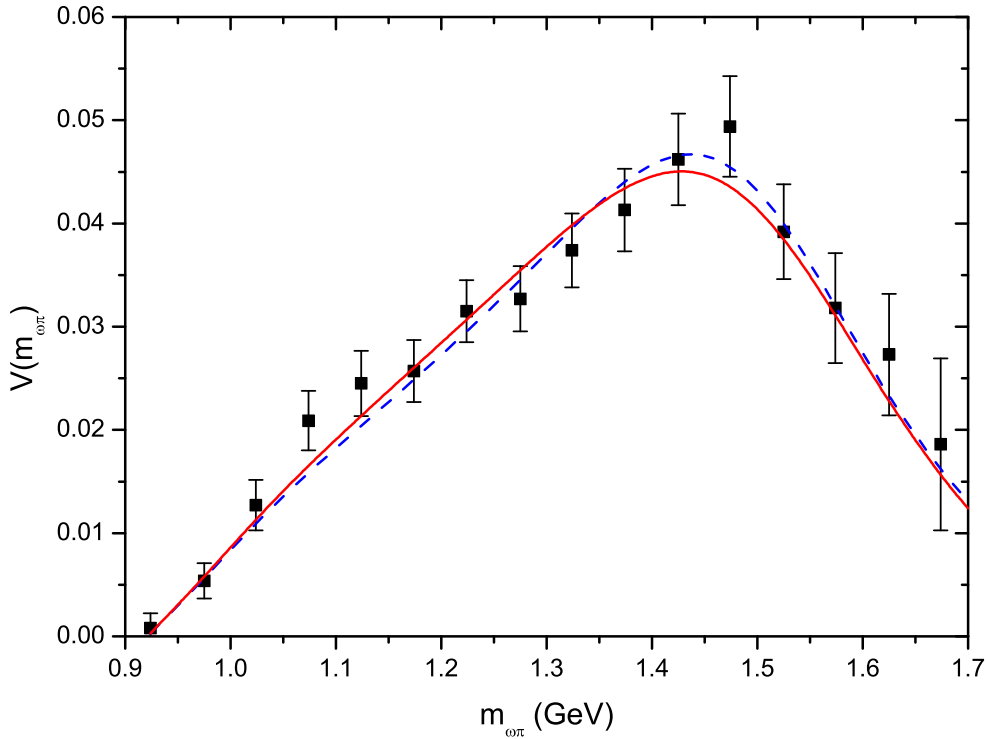}
\caption{ Spectral function for $\tau^- \rightarrow \omega\pi^-
\nu_\tau$. The experimental data are taken from~\cite{CLEO}. }
\label{figVtauOmegapi}
\end{center}
\end{figure*}

The values for the fit parameters are summarized in Table~\ref{tablepar}.
The results of $\Gamma(K^{\ast0}\rightarrow
K^{0}\gamma)$, $\Gamma(K^{\ast+}\rightarrow K^{+}\gamma)$ and
$\frac{\Gamma(K^{\ast0}\rightarrow
K^{0}\gamma)}{\Gamma(K^{\ast+}\rightarrow K^{+}\gamma)}$ from the fits are given
in Table~\ref{tableKstarK}.
The final results corresponding to the $e^{+}e^{-}\rightarrow K^{\ast}(892)K $ processes
and the $\omega\pi$ spectral function from $\tau^- \rightarrow \omega\pi^- \nu_\tau$
are shown in Figs.~\ref{figkstark} and \ref{figVtauOmegapi}, respectively.
In these two figures, we distinguish different results from
Fit~I and Fit~II by solid and dashed lines, respectively.
The resulting curves for the $\omega \rightarrow\pi^0\gamma^*$ form factors are plotted in Fig.~\ref{figffomega}, together
with the results from Refs.~\cite{Leupold10,Leupold12}.

Some comments about the fitting results are given below.

Firstly, we observe that Fit~I is slightly preferred over
Fit~II, according to the values of $\chi^2/$degrees of freedom (d.o.f). Nevertheless, due
to the fact that the values of $\chi^2/d.o.f$ are already bigger
than 2, the preference is not really significant. Regarding to the
somewhat large values of $\chi^2/d.o.f$, we find that they are
mainly contributed by the $\omega\to\pi\gamma^\ast$ form factors
from the NA60 and Lepton-G measurements, especially from the data
points in the energy region around $0.5\sim0.63$~GeV. Therefore we
have tried another kind of fits to exclude the data from these two
measurements. It turns out that the
resulting values of $\chi^2/d.o.f$ decrease to around 1.2 in Fit I
and to 1.9 in Fit II and we do not observe significant changes of
the fit parameters comparing with the results in
Table~\ref{tablepar}. Later we will come back to analyze the
$\omega\to\pi\gamma^\ast$ form factor in detail.

Secondly, regarding to the $c_i$ and $d_i$ parameters given in Table~\ref{tablepar}, we see that the resulting values from the two fits are more or less compatible.
The situation for the masses and widths of broad resonances is different. Roughly speaking, we observe that the determinations of the masses and widths for $\phi'$ from
the two fits are consistent, or at least not so different from each other. This conclusion also holds roughly for the masses of $K^{*'}$.
Nevertheless, we find the masses and widths of $\rho'$ from the two fits are clearly not compatible, indicating that
the mass and width of the $\rho'$ are more sensitive to the forms of energy-dependent widths, compared to the $\phi'$ case.

Focusing on the mass and width of $\phi'$, our determinations in Table~\ref{tablepar} are in good agreement with the
corresponding parameters of $\phi(1680)$ reported in PDG~\cite{Pdg}. Though the mass of $\phi'$ in Ref.~\cite{babar08} is compatible with PDG, its width is somewhat larger.
In addition, the uncertainties in our determination of the mass and width for $\phi'$ are much smaller than
those obtained in Ref.~\cite{babar08}.
Furthermore, we do not need to add the so-called background terms by hand in our amplitude, while this is not the case
in the experimental analyses~\cite{babar08}. All of the terms appearing in our amplitudes
are obtained from the symmetry allowed operators illustrated in Sect.~\ref{sect.lag}, where the contact terms and
contributions from $\rho(770), \omega(782), \phi(1020)$ and $\omega'(1420)$
are incorporated in a consistent way.

Regarding to the $\rho'$ resonance, we should mention that the spectra of the excited $\rho$ resonances are still not
clear in PDG~\cite{Pdg}. The value of the $\rho'$ mass from the 
recent determination of SND collaboration~\cite{SND2013} 
is around 1490~MeV, which is clearly lower than our determinations in Table~\ref{tablepar}. 
As mentioned previously, we find that the resulting $\rho'$ masses from our fits vary accordingly by using different 
forms of energy-dependent widths. In Ref.~\cite{SND2013}, a constant width is used for the $\rho'$ resonance, 
which is obviously different from ours. We have explicitly checked that 
if a constant width for the $\rho'$ is used our determination 
for its mass decreases to around 1500~MeV in both fits, which becomes close to the value in Ref.~\cite{SND2013}.  
The  masses and widths of $\rho'$ from our fits lie between $\rho(1450)$ and $\rho(1700)$. 
Notice that there are roughly two peaks in the experimental isovector component of the $\sigma_{K^{\ast}K}$ cross sections,
see Fig.~\ref{figkstark}(c), which may be due to the contributions of $\rho(1450)$ and $\rho(1700)$.
However since the statistics are still very poor, we only include one set of the $\rho'$ resonances in the current fit.
This indicates that the $\rho'$ here may be regarded as a combined effect of $\rho(1450)$ and $\rho(1700)$.

Finally, we find the mixing angles of $\omega'-\phi'$ from the two fits are roughly consistent. The resulting angle is around $20^{\circ}$ and it is clearly different
from the ideal mixing case. This indicates that the $1/N_C$ suppressed operator in Eq.~\eqref{lagemv1} is quite important in the determination
of the masses of excited vector resonances, since the ideal mixing will result if only the leading order operators at large $N_C$ are included for the resonance masses.

\subsection{Anatomy of the $SU(3)$ breaking mechanism in $K^{\ast}\longrightarrow K \gamma $ decays}

For the ratio $\frac{\Gamma(K^{\ast 0}\rightarrow K^{0}\gamma)}{\Gamma(K^{\ast\pm}\rightarrow K^{\pm}\gamma)}$, the $SU(3)$ symmetry tells us that
$\frac{\Gamma(K^{\ast 0}\rightarrow
K^{0}\gamma)}{\Gamma(K^{\ast+}\rightarrow K^{+}\gamma)}\mid
_{SU(3)}=4$~\cite{Donnell}, while the experimental value is
$\frac{\Gamma(K^{\ast 0}\rightarrow K^{0}\gamma)}{\Gamma(K^{\ast+}\rightarrow K^{+}\gamma)}\mid
_{\rm Ex}\simeq2.3$~\cite{Pdg}.
The $SU(3)$ symmetry breaking effect in the $K^{\ast}\longrightarrow K \gamma $ decay has been discussed in many previous works
~\cite{Durso,Bagan,Bramon,Benayouna}. In the context of vector meson
dominance (VMD) model~\cite{Sakurai}, Ref.~\cite{Durso} points out that
there is a significant contribution from an intermediate
$s\overline{s}$ pair ($\phi$ meson) in these decays, and if all partial $s\overline{s}$ production is suppressed
by only (20-30)\% relative to those involving nonstrange pairs, the ratio of the
neutral to charged $K^{*}$ decay widths would be brought into line
with the data. Including the $SU(3)$ symmetry breaking in the hidden local
symmetry model~\cite{Bramon,Benayouna}, it is found that the rate of $\Gamma(K^{\ast 0}\rightarrow K^{0}\gamma)$ is
reduced comparing with the $SU(3)$ symmetry prediction and
especially it is pointed out that the vector mesons should be ``renormalized'' ~\cite{Benayouna}.

In the framework of R$\chi$T, there is a special operator in Eq.~\eqref{lagvjp}, namely
the $O_{VJP}^4(=\frac{ic_4}{M_V}\epsilon_{\mu\nu\rho\sigma}\bra V^{\mu\nu}[f_-^{\rho\sigma},\chi_+]\ket)$ term, which
breaks the $SU(3)$ symmetry by the factor $(m_s-m_u)$ and exclusively contributes to the charged processes $K^{\ast\pm}\rightarrow
K^{\pm}\gamma$. Hence the value of the coupling $c_4$ is important to decode the $SU(3)$ symmetry mechanism in the present work.

In fact, the $c_4$ value has been determined in Ref.~\cite{Dumm}. In that reference, the authors only included the lightest multiplet of vector resonances,
so the heavier degrees of freedom, such as the excited $\rho'$ state, are implicitly included in their $c_4$.
Through fitting the first 6 data points of the isovector component of
$e^{+}e^{-}\longrightarrow K^{\ast}(892)K\rightarrow K_S K^{\pm}\pi^{\mp} $,  they get $c_4=-0.047 \pm 0.002$ and
through fitting the branching ratios of $\tau\longrightarrow
K\bar{K}\pi\nu_{\tau}$ they obtain $c_4=-0.07 \pm 0.01$. Clearly both values determined in Ref.~\cite{Dumm} are
not compatible with our results in Table~\ref{tablepar} and the magnitudes of $c_4$ in our fits are around one order smaller than those in Ref.~\cite{Dumm}.

We would like to bring the attention that in Ref.~\cite{Guo:2010dv} an unexpected phenomenon is observed by using $c_4=-0.07$ in the radiative decay
of $\tau\to K\gamma \nu_\tau$. Typically one would expect that for a low energy cutoff of the photon in $\tau\to K\gamma \nu_\tau$, the decay
rate should be dominated by the internal bremsstrahlung (IB) contribution. For example, to take the cutoff for the photon energy at 50~MeV, the IB
part contributes more than 90\% of the decay rate for the $\tau\to \pi\gamma \nu_\tau$ process, see Table I in Ref.~\cite{Guo:2010dv}.
However if one looks at Table II in the same reference, the contribution from the IB part in $\tau\to K\gamma \nu_\tau$ dramatically decreases
to 6\% by taking $c_4=-0.07$ and in this case the $c_4$ term dominates the whole process. In the case with $c_4=0$, one can see some reasonable
results. This clearly gives us a hint that the large magnitude of $c_4$ (compared to 0.07), does not lead to reasonable results in the radiative
tau decays. Therefore our current determination for $c_4$, around one order smaller in magnitude than the values given in Ref.~\cite{Dumm},
seems more meaningful. However due to the lack of experimental measurements on the radiative tau decays,
we can still not make a concrete conclusion whether the $c_4$ values in Table~\ref{tablepar} will reasonably
describe the $\tau\to K\gamma \nu_\tau$ process.
A future measurement on this channel is definitively helpful to answer this question.

In order to have a better understanding of the origin of the differences for the $c_4$ values between ours and Ref.~\cite{Dumm}, we next employ the same
$VVP$ Lagrangian as used in Ref.~\cite{Dumm}, indicating in the discussions below we exclude the excited vector resonances in our theory.
In this case, it is interesting to point out that after taking into account the high energy constraints given in the previous work~\cite{chen,Femenia}
\begin{eqnarray}
4c_3+c_1&=&0,\nonumber\\ c_1-c_2+c_5&=&0,\nonumber\\
c_5-c_6&=&\frac{N_C}{64\pi^2}\frac{M_V}{\sqrt{2}F_V},\nonumber\\
d_1+8d_2-d_3&=&\frac{F^2}{8F_V^2},\nonumber\\
d_3&=&-\frac{N_C}{64\pi^2}\frac{M_V^2}{F_V^2}\label{eqconstraint2}\,,
\end{eqnarray}
we can totally predict the $K^{\ast 0}\rightarrow K^{0}\gamma$ decay width, which turns out to be $109.1$~KeV and agrees well with
the experimental value $(116.2\pm11.6)$~KeV~\cite{Pdg}.
For the charged process $K^{\ast+}\rightarrow K^{+}\gamma$, we only have one free parameter, $c_4$, to determine its decay width.
Therefore, one could use the experimental information $\Gamma_{K^{*+}\rightarrow K^+\gamma}=(50.3\pm 5.5)$~KeV to determine the value of $c_4$.
We then obtain two solutions
\begin{eqnarray}\label{c4sol}
c_4=0.0003\pm 0.0007\,\,\,  {\rm or} \,\, -0.0251\pm0.0007\,.
\end{eqnarray}

By setting $c_4=-0.0251\pm 0.0007$ in the decay amplitudes, we find the $c_4$ term then dominates the $K^{*+}\rightarrow K^+\gamma$ process,
which indicates that the $SU(3)$ symmetry breaking effect overwhelmingly controls this process and clearly contradicts
a general rule that this effect should be at most around 30\%. This tells us only the solution of $c_4=0.0003\pm 0.0007$ in Eq.~\eqref{c4sol}
corresponds to the physical one. Indeed this observation also confirms the conclusion about the role of
$c_4$ in radiative tau decays~\cite{Guo:2010dv}, where a small magnitude of $c_4$ is preferred.
In the determination of Eq.~\eqref{c4sol}, we do not explicitly include the vector resonance excitations and the corresponding results
after the incorporation of the excited states can be found in Table~\ref{tablepar},
which magnitude of $c_4$ is still one order smaller than that in Ref.~\cite{Dumm}.
In the present work we provide another important and easy way to determine the $c_4$ value, comparing with Ref.~\cite{Dumm},
which can be useful for the future study of various tau decays.

In order to have a clear idea about how different parts contribute to the $SU(3)$ breaking,
we decompose the amplitudes $g_{K^{\ast0}K^{0}\gamma}$ and $g_{K^{\ast+}K^{+}\gamma}$ in
Eqs.\eqref{eqKstar0K} and \eqref{eqKstarpK} into the two parts: the $SU(3)$ symmetric term and  $SU(3)$ breaking term.
The decomposition takes the form
\begin{eqnarray}
&& g_{K^{\ast0}K^{0}\gamma}^{SU(3) symmetry}= -\frac{4\sqrt{2}}{3F_K
M_V M_{K^*}} \bigg[(c_1+c_2+8c_3-c_5)m_K^2 +(c_2+c_5 -c_1-2c_6)M_{K^*}^2 \bigg]
\nonumber\\
&&-\frac{2 F_V}{3F_K M_{K^*}} \bigg( \frac{1}{M_V^2}
-\frac{3}{M_V^2}-\frac{2}{M_V^2} \bigg)\bigg[( d_1+8d_2-d_3)m_K^2  +d_3
M_{K^*}^2 \bigg]\nonumber\\&&
-\frac{F_{V_1}}{3F_K M_{K^*}}
\bigg(\frac{1}{M_{V_1}^2} -\frac{3}{M_{V_1}^2}-\frac{2}{M_{V_1}^2} \bigg)
(d_m m_K^2+d_M M_{K^*}^2), \nonumber\\
&& g_{K^{\ast0}K^{0}\gamma}^{SU(3) breaking}= -\frac{2 F_V}{3F_K
M_{K^*}} \bigg[ \bigg( \frac{1}{M_\omega^2}
-\frac{3}{M_\rho^2}-\frac{2}{M_\phi^2} \bigg)
\nonumber \\&&
- \bigg( \frac{1}{M_V^2}
-\frac{3}{M_V^2}-\frac{2}{M_V^2} \bigg) \bigg]
\bigg[( d_1+8d_2-d_3)m_K^2 +d_3
M_{K^*}^2 \bigg] \nonumber\\&&
-\frac{F_{V_1}}{3F_K M_{K^*}} \bigg\{ \bigg[
\frac{(2\sqrt{2}\cos\theta_V-\sin\theta_V)\sin\theta_V}{M_{\omega'}^2}
-\frac{3}{M_{\rho'}^2}
-\frac{(\cos\theta_V+2\sqrt{2}\sin\theta_V)\cos\theta_V}{M_{\phi'}^2}
\bigg]
\nonumber\\&& -\bigg(\frac{1}{M_{V_1}^2}
-\frac{3}{M_{V_1}^2}-\frac{2}{M_{V_1}^2}  \bigg) \bigg\} (d_m m_K^2+d_M
M_{K^*}^2)\,, \label{eqKstar0KSU3symPlusbreaking}\end{eqnarray}
\begin{eqnarray}
&&g_{K^{\ast+}K^{+}\gamma}^{SU(3) symmetry}= \frac{2\sqrt{2}}{3F_K
M_V M_{K^*}} \bigg[ (c_1+c_2+8c_3-c_5)m_K^2 +(c_2+c_5 -c_1-2c_6)M_{K^*}^2 \bigg]
\nonumber\\
&&-\frac{2 F_V}{3F_K M_{K^*}} \bigg( \frac{1}{M_V^2}
+\frac{3}{M_V^2}-\frac{2}{M_V^2} \bigg) \bigg[( d_1+8d_2-d_3)m_K^2  +d_3
M_{K^*}^2\bigg]\nonumber\\&& -\frac{F_{V_1}}{3F_K M_{K^*}}
\bigg( \frac{1}{M_{V_1}^2} +\frac{3}{M_{V_1}^2}-\frac{2}{M_{V_1}^2} \bigg) (d_m
m_K^2+d_M M_{K^*}^2), \nonumber\\
&& g_{K^{\ast+}K^{+}\gamma}^{SU(3) breaking}= -\frac{2 F_V}{3F_K
M_{K^*}} \bigg[ \bigg( \frac{1}{M_\omega^2}
+\frac{3}{M_\rho^2}-\frac{2}{M_\phi^2} \bigg)
\nonumber \\ &&
-\bigg( \frac{1}{M_V^2}
+\frac{3}{M_V^2}-\frac{2}{M_V^2} \bigg) \bigg] [( d_1+8d_2-d_3)m_K^2 +d_3
M_{K^*}^2]\nonumber\\&&
-\frac{F_{V_1}}{3F_K M_{K^*}} \bigg\{ \bigg[
\frac{(2\sqrt{2}\cos\theta_V-\sin\theta_V)\sin\theta_V}{M_{\omega'}^2}
+\frac{3}{M_{\rho'}^2}
-\frac{(\cos\theta_V+2\sqrt{2}\sin\theta_V)\cos\theta_V}{M_{\phi'}^2}
\bigg]\nonumber\\&& -\bigg(\frac{1}{M_{V_1}^2}
+\frac{3}{M_{V_1}^2}-\frac{2}{M_{V_1}^2} \bigg) \bigg\} (d_m m_K^2+d_M
M_{K^*}^2)+\frac{16\sqrt{2}}{3F_K M_V M_{K^*}}c_4(m_K^2-m_\pi^2)\,.
\label{eqKstarPKSU3symPlusbreaking}\end{eqnarray}
The criteria to make the above decomposition is that in the $SU(3)$ symmetric terms we do not distinguish
the mass differences of intermediate vectors and also assume the ideal mixing for $\omega'-\phi'$. So it is easy to check that
$|g_{K^{\ast0}K^{0}\gamma}^{SU(3) symmetry}/g_{K^{\ast+}K^{+}\gamma}^{SU(3) symmetry}|=2$, consistent with the
$SU(3)$ symmetry requirement.
While for the $SU(3)$ breaking parts, there are three sources: different masses of the intermediate vector resonances,
the non-ideal mixing angle of the $\omega'-\phi'$ and the $c_4$ term in the charged process.
The mass differences and the mixing angle of $\omega'-\phi'$ are directly related to the resonance operators in Eqs.~\eqref{lagemr} and \eqref{lagemv1}.
Therefore it is crucial to include these mass splitting parameters in the Lagrangian approach in order to systematically study the $SU(3)$ symmetry mechanism.

Next let us  take the results from Fit~I in Table~\ref{tablepar} to make a concrete analysis on the
strengths of different parts contributing to the $SU(3)$ symmetry breaking.
The following numbers can be straightforwardly obtained
according to the parameters from Fit I in Table~\ref{tablepar}
\begin{eqnarray}
&&\frac{g_{K^{\ast0}K^{0}\gamma} }{g_{K^{\ast+}K^{+}\gamma}
}=\frac{g_{K^{\ast0}K^{0}\gamma}^{SU(3)symmetry
}+g_{K^{\ast0}K^{0}\gamma}^{SU(3)breaking}}{g_{K^{\ast+}K^{+}\gamma}^{SU(3)symmetry
}+(g_{K^{\ast+}K^{+}\gamma}^{SU(3)breaking}-g_{K^{\ast+}K^{+}\gamma}^{c_4
})+g_{K^{\ast+}K^{+}\gamma}^{c_4
}}\nonumber\\
&& =\frac{-16.1+2.5 }{8.0+2.0-1.6 }\ .
\end{eqnarray}
It is obvious that these $SU(3)$ breaking effects suppress the neutral $K^\ast$ decay width but
enhance the charged $K^\ast$ mode, as a result the ratio between
them becomes compatible with the data. 
Notice that among the $SU(3)$ breaking effects in $K^{\ast+}\rightarrow K^{+}\gamma$, the
$c_4$ term contributes at the same order as the other $SU(3)$
breaking effects in R$\chi$T, although the values of the $c_4$ parameter, with the order of $10^{-3}$ in 
Table~\ref{tablepar}, seem unnaturally small. The small magnitude of $c_4$ can be attributed to the 
normalization of the $VJP$ operators in Eq.~\eqref{lagvjp}.  This conclusion can be easily verified by comparing  
the values of other $VJP$ couplings with $c_4$. 
We notice that one $VJP$ operator is considered in Ref.~\cite{Leupold10}, 
namely the $e_A$ term in Eq.(15) of this reference. It is easy to obtain that the $e_A$ term is related to our $c_6$ term
in Eq.~\eqref{lagvjp} through $c_6= e_A/(8\sqrt2 e)$, 
with $e_A$ estimated to be around $1.5\times 10^{-2}$ in Ref.~\cite{Leupold10}, which leads to 
$c_6 \simeq 4.4 \times 10^{-3}$. Therefore we show that the magnitude of $c_4$ parameter given in Table~\ref{tablepar} 
is at the same order as other $c_i$ couplings in the $VJP$ Lagrangian~\eqref{lagvjp} estimated from other works.

\subsection{ Discussion on the $\omega\longrightarrow \pi^0 \gamma^* $ transition form factor}

Recently, the new measurements on the $\omega\rightarrow \pi^0\gamma^*$ form factors from the NA60 collaboration~\cite{NA6011,NA6009} have
activated several updated theoretical works~\cite{Leupold10,Leupold12,Kubis}. The main problem behind is that the conventional
VMD approach can not well incorporate these data, so new ideas are proposed to have a deeper understanding why the successful
VMD is not working here.

In Refs.~\cite{Leupold10,Leupold12}, a new counting scheme different from
R$\chi$T is used: the masses of both vector mesons and pseudoscalar
mesons are treated as small expansion parameters.
So in their counting scheme, the Lagrangian that accounts for the decay mechanism through exchanging an intermediate vector
meson belongs to the leading order, while the Lagrangian responsible for the contact terms in the decay amplitude
belongs to the next-to-leading-order.
In Ref.~\cite{Leupold10}, all of the leading order contributions in their scheme and some incomplete parts of
next-to-leading-order contributions are considered in their $\omega\rightarrow \pi^0 \gamma^* $ form factor.
After a careful check, we realize that the $VVP$-type of Lagrangian in Eq.~\eqref{lagvvp} involving
the lowest vector multiplet coincides with the so-called leading order Lagrangian proposed in Ref.~\cite{Leupold10}
up to some normalization factors. For example, from Eqs.(10) and (15) of Ref.~\cite{Leupold10}, it is easy to see that their $b_A$
corresponds to our $d_2$, their $h_A$ relates with our $d_1+d_3$
\footnote{Using the Shouten identity
$g_{\sigma\rho}\epsilon_{\alpha\beta\mu\nu}+g_{\sigma\alpha}\epsilon_{\beta\mu\nu\rho}
+g_{\sigma\beta}\epsilon_{\mu\nu\rho\alpha}+g_{\sigma\mu}\epsilon_{\nu\rho\alpha\beta}
+g_{\sigma\nu}\epsilon_{\rho\alpha\beta\mu}=0$
and the equation of motion $\nabla^\mu
u_\mu=\frac{i}{2}(\chi_{-}-\frac{1}{N_F}\langle\chi_{-}\rangle)$~\cite{Cirigliano},
we have
$d_{1}\epsilon_{\mu\nu\rho\sigma}\langle\{V^{\mu\nu},V^{\rho\alpha}\}\nabla_\alpha
u^\sigma\rangle=\frac{1}{4}d_{1}\epsilon_{\mu\nu\rho\sigma}\langle\{V^{\mu\nu},V^{\rho\sigma}\}\nabla_\alpha
u^\alpha\rangle=\frac{i}{8}d_{1}\epsilon_{\mu\nu\rho\sigma}\langle\{V^{\mu\nu},V^{\rho\sigma}\}\chi_-\rangle
-\frac{i}{12}d_{1}\epsilon_{\mu\nu\rho\sigma}\langle
V^{\mu\nu}V^{\rho\sigma}\rangle\langle\chi_-\rangle$.
The last term with two traces can be neglected, as it is $1/N_C$ suppressed. Since the parameters
of $d_1$ and $d_2$ always appear in the combination as $d_1+8d_2$ in our case,
the number of independent terms with two vector resonances in this work agrees with that used in Ref.~\cite{Leupold10}. }.
The $e_A$ term in Ref.~\cite{Leupold10}, which belongs to the next-to-leading order in their counting,
is equivalent to our $c_6$ term, one of the six $VJP$-type operators in Eq.~\eqref{lagvjp}.
One should notice that the authors in Ref.~\cite{Leupold10} mostly focus on the leading order computation and in order to
make a rough quantitative estimate on the next-to-leading order effects they select a particular operator to perform the numerical discussion.
While in our case, the contact terms, i.e. the $VJP$
operators in Eq.~\eqref{lagvjp}, are not suppressed by any counting rule and fully included in the analyses.
Also the excited vector resonances are not considered in Refs.~\cite{Leupold10,Leupold12}.

Though we have some similar operators from the beginning, the QCD high energy
behavior is not appreciated in the discussion of Refs.~\cite{Leupold10,Leupold12}. As we have mentioned in the Introduction,
the transition form factor with proper high energy constraint can be directly applied in other physical quantities, such
as the light-by-light scattering. So we consider it is an improvement to impose the QCD short distance constraints in the form factor.

\begin{figure*}[ht]
\begin{center}
\includegraphics[width=1.0\textwidth]{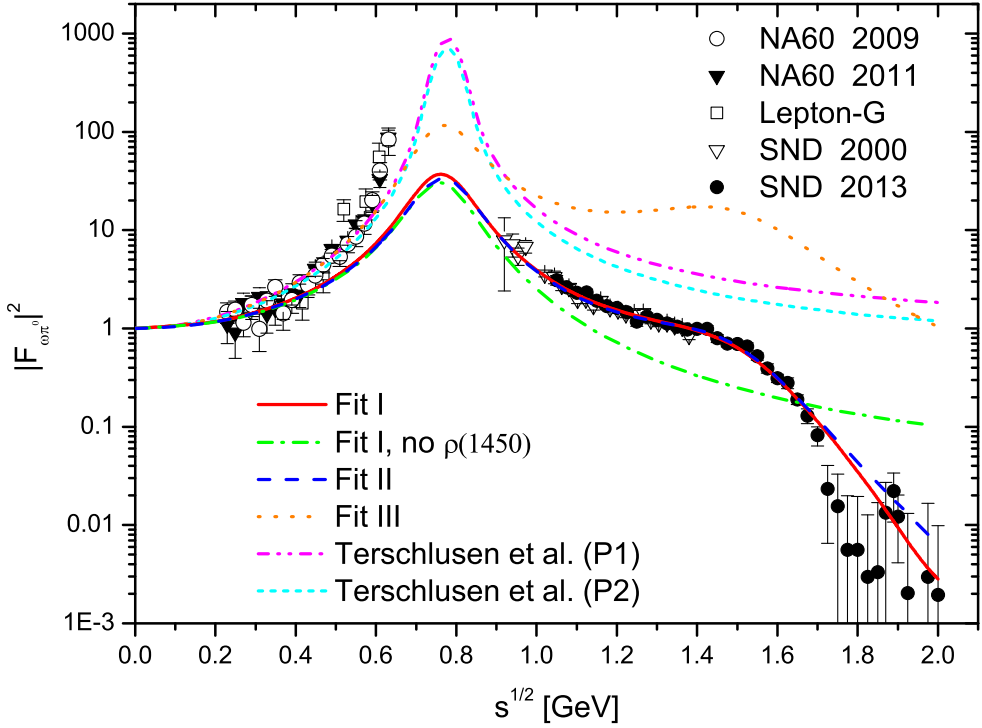}
\caption{  The form factors of $\omega\pi^0\gamma^*$: the
data in the left side of $0.8$~GeV correspond to the $\omega$-decay
processes and those in the right side correspond to the
$\omega$-production processes.  The red solid line denotes the
result from Fit~I and the blue dashed line denotes the result from
Fit~II. The green dash-dotted line denotes the prediction of Fit I
by simply excluding the contributions from the excited $\rho'$
resonance. The orange doted line corresponds to the results of
Fit III in which the $\omega\pi$ form-factor data from SND Collaboration and the $\omega\pi$ spectral function data from tau decays are not included in the analysis. The
magenta dash-dot-dotted and cyan short dashed lines correspond to
the predictions of the parameter set 1 (P1) and parameter set 2 (P2) in
Ref.~\cite{Leupold10}. Sources of the different experimental data
are: open circles~\cite{NA6009}, solid triangles~\cite{NA6011}, open
squares~\cite{LG}, open triangles~\cite{SND2000}, solid
circles~\cite{SND2013}. } \label{figffomega}
\end{center}
\end{figure*}

Next we make a close comparison with Ref.~\cite{Leupold10,Leupold12}. A similar theoretical framework is
employed in the former references as ours, but different experimental data are analyzed. The focus of Refs.~\cite{Leupold10,Leupold12} is the $\omega \to
\pi^0 \gamma^*$ form factors from the $\omega$ decay process, not
from the $\omega$ production. It turns out that the form factors in
Refs.~\cite{Leupold10,Leupold12} can fairly well describe the
$\omega \to \pi^0 \gamma^*$ data from NA60 collaboration. Then it is
interesting for us to perform another fit to only consider the
same type of data as in Refs.~\cite{Leupold10,Leupold12}, which will
be referred as Fit III. To be more specific, we only include the
form-factor data from the $\omega$ decay \cite{LG,NA6009,NA6011} in
Fit III and exclude these from the $\omega$
production~\cite{SND2000,SND2013} and the $\tau^- \rightarrow
\omega\pi^- \nu_\tau$ spectral function~\cite{CLEO}. As shown in
Fig.~\ref{figffomega}, the final output of our Fit III (the orange
dotted line) tends to be quite similar with those from
Ref.~\cite{Leupold10} in the focused energy region, which is below
the $\omega$ mass. However, we observe that the $\omega\pi$ form
factors in the production region are clearly incompatible with the
data from SND collaboration taking the resulting parameters from Fit
III, see the orange dotted lines in the region above 0.9~GeV in Fig.~\ref{figffomega}.
Similarly, we find that the form factors in
Refs.~\cite{Leupold10,Leupold12}, when extrapolated to the
production region~\footnote{ The plots in the energy region above
$M_\omega-m_\pi$ are not explicitly given in
Refs.~\cite{Leupold10,Leupold12}. We have exploited the theoretical
formulas presented in t references to plot the curves in
Fig.~\ref{figffomega}.}, are in contradiction with the SND data as
well, see the magenta dot-dot-dashed and the cyan dashed lines in Fig.~\ref{figffomega}. This tells us that a good reproduction of the form factors in
$\omega$ decay does not necessarily guarantee a consistent
description of the form factors in $\omega$ production. Therefore it
is important to consider both types of data simultaneously, as done
in our Fit I and Fit II.
After including the SND and CLEO data in the analysis, i.e. the fits presented in Table~\ref{tablepar}, we find that our description
of the NA60 data becomes obviously worse, see the difference between red solid line and the orange dotted line in Fig.~\ref{figffomega}.
It turns out that our theoretical framework starts to be incompatible with the NA60 data above 0.5~GeV.

In this work the excited vectors are explicitly included, which are obviously important in the $\omega$ production processes from the SND~\cite{SND2000,SND2013}
and CLEO~\cite{CLEO} measurements on the $\omega\pi$ form factors.
While, it is also interesting to see to what extent the excited vector resonances can affect
the $\omega\pi$ form factor in the $\omega$ decay, i.e. the energy region below $M_\omega-m_\pi$. In Fig.~\ref{figffomega}, we
designate the green dot-dashed line to the situation in which we simply exclude the $\rho'$ contribution.
It is obvious that the $\rho'$ plays an important role in describing the SND Collaboration data,
while its effect in the low energy region is quite small.

In a short summary about the fit to $\omega\pi$ form factor, we could describe the NA60 data in our case to the same extent
as in Ref.~\cite{Leupold10,Leupold12}, if we had not considered the SND and CLEO data.
After the inclusion of the latter two measurements, we conclude that our description is still
inadequate for the $\omega\to\pi\gamma^*$ form factors
from the NA60 measurements in the energy region near the kinematical boundary, i.e. $M_\omega-m_\pi$.
Hence we consider this is still an open problem in our framework.

\indent

\section{Conclusions}
\label{conclu}

In this work, the resonance chiral theory is used to study
$K^{\ast}(892)\rightarrow K\gamma$, $e^{+}e^{-}\rightarrow
K^{\ast}(892)K $ processes, the $\omega\pi^0 \gamma^* $ transition
form factor, and the spectral function for $\tau^- \rightarrow
\omega\pi^- \nu_\tau$. Through fitting the corresponding
experimental data, the values of resonance couplings, that can not
be fixed through the QCD high energy constraints, are then provided.
The masses and widths of our $\phi'$ resonance are quite compatible with
the $\phi(1680)$ in PDG~\cite{Pdg}. While our $\rho'$ looks more
like a combined effect of the proposed $\rho(1450)$ and $\rho(1700)$
from PDG. The mass and width of the $\rho'$ resonance are found to be sensitive to the forms of the energy-dependent widths used in the propagator, while
the parameters for the $\phi'$ are more or less stable.
The resulting values of the $\omega'-\phi'$ mixing angle are found to be around
$20^{\circ}$,  clearly different from the ideal mixing case as in the $\omega-\phi$ case.
The resonance coupling $c_4$ obtained here is around one order smaller in magnitude than
those in literature and our results seem more meaningful when
considering the radiative tau decays. With the $c_4$ value obtained
here, we analyze the different $SU(3)$ breaking effects contributing
to the ratio of $\frac{\Gamma(K^{\ast 0}\rightarrow
K^{0}\gamma)}{\Gamma(K^{\ast\pm}\rightarrow K^{\pm}\gamma)}$ in
detail. And we find that the three sources of $SU(3)$ symmetry breaking effects, e.g. the different masses
of intermediate resonances, the non-ideal mixing angle of $\omega'-\phi'$ and non-vanishing value of $c_4$,
are more or less equally important at the numerical leve.

The excited vector resonances are found to be essential to reproduce the $\gamma^*\to\omega\pi^0 $ form-factor data from SND
and the CLEO $\omega\pi$ spectral function from tau decays.
Although the low energy $\omega\pi$ form-factor data from NA60 can be well reproduced in our approach,
the steep rise behavior close to the upper kinematical boundary region $\sqrt{s}\leq M_\omega-m_\pi$, is
still not fully understood in the current work, even after taking into account the excited vector resonances.

\begin{center}
\section*{Acknowledgments}
\end{center}
This work is supported in part by  National Nature Science
Foundations of China (NSFC) under contract Nos. 10925522, 11021092, 11105038 and 11035006.
YHC acknowledges support by Grants No.11261130311
(CRC110 by DFG and NSFC).
ZHG acknowledges the grants from Education Department of Hebei Province with contract number YQ2014034, Natural Science
Foundation of Hebei Province with contract number A2011205093,
Doctor Foundation of Hebei Normal University with contract number
L2010B04.

\end{document}